%% file: main.tex
\documentclass{article}
\usepackage{spconf,amsmath,graphicx}
\usepackage{extra_styles}

\usepackage{cite}
\usepackage{amssymb,amsfonts}
\usepackage{algorithmic}
\usepackage{graphicx}
\usepackage{textcomp}
\usepackage{xcolor}
\usepackage{booktabs}
\usepackage{algorithm}
\usepackage{array}
\usepackage{subcaption}
\usepackage{setspace}

\usepackage{mathtools}
\usepackage{url}
\usepackage[breaklinks, colorlinks, pagebackref]{hyperref}
\usepackage[subtle,mathspacing=normal]{savetrees}

\title{SCORE: Scaling audio generation using Standardized COmposite REwards}

\name{
    \begin{tabular}{c}
        Jaemin Jung$^{1*}$, Jaehun Kim$^{1*}$, Inkyu Shin$^{2}$, Joon Son Chung$^{1}$ \thanks{\hspace{-10pt}* These authors contributed equally.}
    \end{tabular}
    }

\address{$^{1}$Korea Advanced Institute of Science and Technology, South Korea\quad
$^{2}$ByteDance Seed, USA
}

\begin{document}
\maketitle
\begin{abstract}

The goal of this paper is to enhance Text-to-Audio generation at inference, focusing on generating realistic audio that precisely aligns with text prompts.
Despite the rapid advancements, existing models often fail to achieve a reliable balance between perceptual quality and textual alignment.
To address this, we adopt Inference-Time Scaling, a training-free method that improves performance by increasing inference computation.
We establish its unexplored application to audio generation and propose a novel multi-reward guidance that equally signifies each component essential in perception.
By normalizing each reward value into a common scale and combining them with a weighted summation, the method not only enforces stable guidance but also enables explicit control to reach desired aspects.
Moreover, we introduce a new audio-text alignment metric using an audio language model for more robust evaluation.
Empirically, our method improves both semantic alignment and perceptual quality, significantly outperforming naive generation and existing reward guidance techniques.
Synthesized samples are available on our demo page: \url{https://mm.kaist.ac.kr/projects/score/}

\end{abstract}
\begin{keywords}
text-to-audio generation, inference-time scaling, diffusion model
\end{keywords}
\def\ours{AWARD}
\input{sec/1_introduction}

\input{sec/2_method}
\input{sec/3_experiment}
\input{sec/4_results}

\input{sec/5_conclusion}

\clearpage
\ninept

\setstretch{0.85}

\bibliographystyle{IEEEbib}
\bibliography{shortstrings,refs}

\end{document}

%% file: sec/1_introduction.tex
\section{Introduction}
\input{figs/04_main}

The growing demand for realistic audio in media like movies, audiobooks, and games has accelerated progress in Text-to-Audio (T2A) generation.
The objective of T2A is to synthesize high-quality audio that accurately matches text~\cite{audiogen}.
Recent advancements, particularly with the diffusion mechanism~\cite{ho2020denoising, song2020denoising}, have enabled the creation of rich, varied soundscapes~\cite{audioldm, audioldm2, evans2024fast, hai2024ezaudio, jung2025voicedit, huang2023make, ghosal2023text}. 
Stable Audio~\cite{evans2024fast} incorporates a latent diffusion model with text and onset conditioning mechanism that generates high-quality audio of variable length. 
EZAudio~\cite{hai2024ezaudio} integrates Diffusion Transformer and classifier-free-guidance rescaling that improve controllability and output diversity.

However, even advanced T2A models often fail to reliably satisfy the two key aspects: audio-text alignment and perceptual quality. This issue is particularly pronounced for diffusion models due to their stochastic nature in the generation process.
Diffusion-based T2A models generate audio by iteratively denoising a random noise vector, resulting in high output variance for a single prompt.
This variance, while beneficial for creative diversity, significantly hinders the ability to reliably generate a specific, desired sound.

A promising, training-free solution is Inference-Time Scaling (ITS), which allocates additional computation at inference to guide the generation towards superior outputs, such as higher sample quality and alignment with text~\cite{its_diffusion}.
Many approaches have introduced different ITS techniques, including Best-of-N (BON) selection~\cite{xie2025sana, lifshitzmulti}, reward-guided particle sampling~\cite{particlesampling_video}, and evolutionary search (EvoSearch)~\cite{evosearch}, all of which effectively improve the output of a naive generation with increased inference computation.
However, while the method was first explored in large language models~\cite{particleLLM, brown2024large, snell2024scaling, wu2024inference} and extended to image and video domains~\cite{reflectdit, particlesampling_video, xie2025sana, singhal2025general, liu2025video}, its application to audio generation is yet to be discovered.
Moreover, most methods provide limited advancements in reward guidance, a crucial component that guides the overall ITS process.
Most existing methods leverage a single reward model as guidance which introduces ``verifier hacking'', a decrease in performance due to over-reliance on a single objective~\cite{its_diffusion}.
Others utilize an aggregation of multiple rewards by converting different values into ranks~\cite{its_diffusion, its_audiosr}, but treating criteria with equal importance under-utilizes the complementary strengths of distinct rewards.

In this paper, we pioneer the application of ITS to T2A generation and conduct comprehensive empirical study of different methods in this domain.
Furthermore, to address the limitations of prior techniques that rely on single rewards or rank aggregation, we propose Standardized Composite Reward (SCORE), a novel reward guidance designed to effectively configure and balance two rewards specifically for T2A tasks: audio quality and audio-text alignment.
Specifically, we normalize each reward with zero mean and unit variance to establish equal influence, and combine them with weighted summation where the weight can be controlled to effectively steer the generation to achieve desired aspect.
Moreover, we assess the text-audio alignment beyond the scope of the reward model by introducing AQAScore, an evaluation metric based on audio language model Audio Flamingo 3~\cite{af3}.
Motivated by VQAScore~\cite{vqascore}, we ask the model with the generated audio and its text prompt and obtain the probability of producing ``yes'' token, serving as a quantitative measure of alignment.

Our experiments demonstrate that SCORE significantly enhances the generation quality of T2A model compared to naive generation, improving both CLAP and PQ metric by $11\%$.
We further verify that SCORE effectively avoids ``verifier hacking" and outperforms existing rank aggregation method in both audio-text alignment and audio quality.
Finally, we show that the generation quality faithfully reflects the adjustment of combination weights, offering a principled approach to prioritize specific attributes of interest.

%% file: figs/04_main.tex
\begin{figure*}[t]
\label{main_fig}
  \centering
  \includegraphics[width=0.98\linewidth]{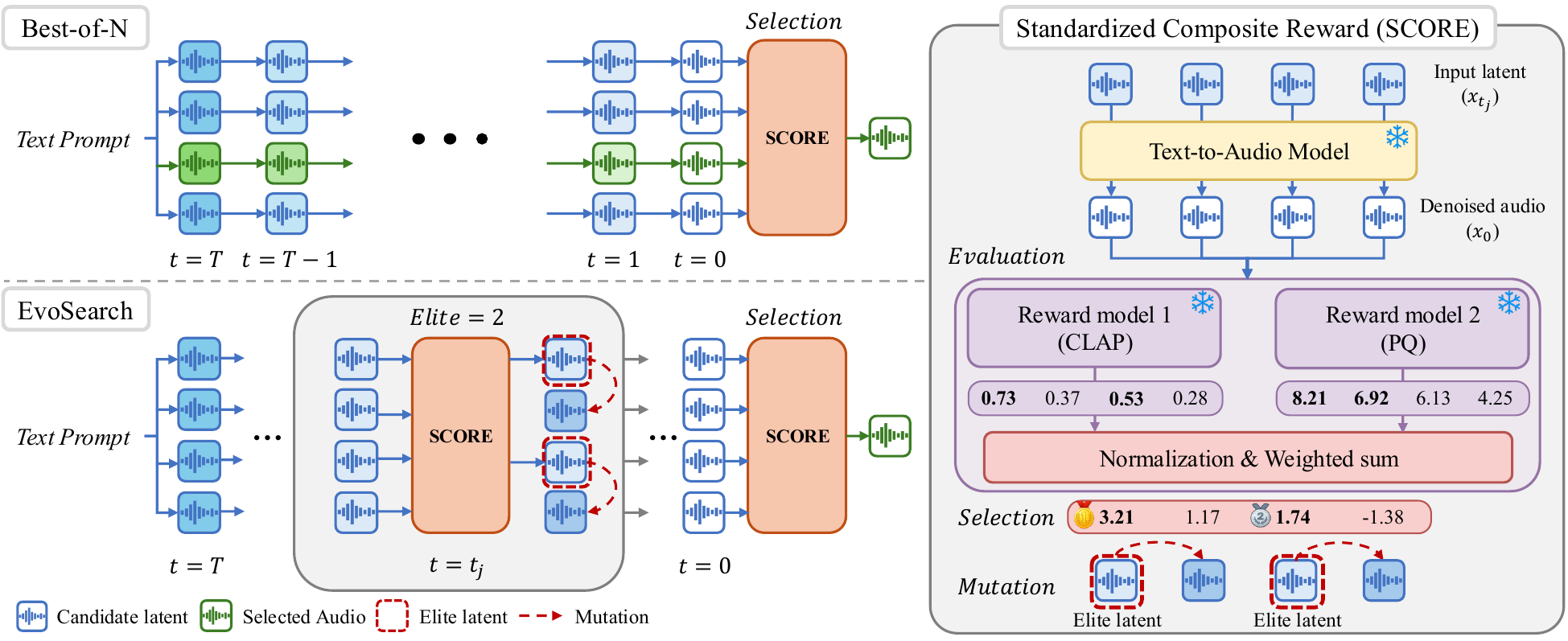}
  \caption{
  \textbf{Overview of our Inference-Time Scaling framework for Text-to-Audio generation.}}
  \label{fig:main}
  \vspace{-2mm}
\end{figure*}

%% file: sec/2_method.tex
\section{Method}
\input{tables/1_main_10}

\subsection{Inference-Time Scaling}
We augment T2A generation with established ITS methods: BON~\cite{its_diffusion} and EvoSearch~\cite{evosearch}.
Both methods initialize multiple noise latents to denoise, and select the best performing sample from the finally produced audios.
To successfully evaluate the samples, they utilize pretrained audio evaluation models as reward models and obtain the reward values.
As shown in \Fref{fig:main}, while BON applies reward guidance only at the end, EvoSearch enforces additional guidance the intermediate denoising stages.
Specifically, when diffusion timesteps lie in predefined evolution steps $\mathcal{T}_{evo}=\{t_1, \dots, t_j, \dots, t_k\}$, the intermediate latents are decoded to audio and fed to reward models for evaluation.
Based on the obtained rewards, high-performing ``elite'' latents are selected and copied with a slight mutation to replace the rejected ones.
This mutation involves adding subtle noise to the latents, increasing the diversity while preserving the core quality.
The evolution process (evaluation, selection, and mutation) occurs in every evolution step in $\mathcal{T}_{evo}$ to iteratively refine the latent candidates, and the final best audio is selected identical to BON.
As EvoSearch utilizes reward models more frequently and applies guidance in the middle of denoising, the method can provide more distinct effect of reward guidance than BON sampling.

\input{figs/01_distribution_2}

\subsection{Multi Reward Models}\label{2.2}

As mentioned earlier, guiding inference with a single reward leads to “verifier hacking," causing the generation to over-adapt to the reward and fail to achieve high overall quality~\cite{its_diffusion}.
To this end, the objective of the proposed method covers two aspects of audio generation: audio-text alignment and high-quality generation.
We select the CLAP~\cite{clap} score as a reward for audio-text alignment, and PQ from Audiobox-Aesthetics~\cite{audiobox_aesthetics} for general audio quality.
However, utilizing multiple reward models poses a critical challenge, because each signal originates from a unique distribution.
For instance, the CLAP score ranges from 0 to 1, whereas the PQ score ranges from 0 to 10, with a high concentration of values around 5 to 7~\cite{audiobox_aesthetics}.
This mismatch in distribution can easily bias the reward guidance.
Therefore, we normalize each distribution to provide equal influence for each reward, which is further elaborated in the next section.

\subsection{Standardized COmposite REward (SCORE)}
Existing methods mitigate imbalances across different scales to minimize bias by converting the reward values into ranks~\cite{its_diffusion, jin2025inference} or by applying a weighted summation~\cite{eyring2024reno, kim2025test}.
In this work, we aim to eliminate the need for a heuristic weight search while utilizing the full-precision values of the rewards, rather than their ranks.
As illustrated in Fig~\ref{fig:distribution}, even when different reward scores are normalized to the same range, their underlying distributions can differ significantly.
This distributional mismatch can disturb the combined reward guidance, necessitating a heuristic search for appropriate weights to counteract this effect.

To address this, we propose Standardized Composite Reward (SCORE), a method that combines each reward signal by normalizing to a zero mean and unit variance, followed by a weighted summation.
This process ensures that multiple reward values can be combined without bias from disparate scales or distributions, and the weight parameter enables the controllability to shift the direction of generation.
Specifically, we first obtain the raw reward scores using each reward model at evolution timestep $t_i$
Following EvoSearch, raw reward score is defined as:
\begin{equation}
s(x_{t_i}) = E_{x_0 \sim p_0(x_0|x_{t_i})} [r(x_0)|x_{t_i}],
\end{equation}
where $x_0$ is the denoised audio from intermediate latent $x_{t_i}$ and $r(\cdot)$ is the reward function.
Next, we normalize the raw score using the mean ($\mu$) and standard deviation ($\sigma$) calculated from the training dataset:
\begin{equation}
z(x_{t_i})=\frac{s(x_{t_i})-\mu}{\sigma}.
\end{equation}
The normalized scores are then combined via a weighted sum to produce the final reward score for the intermediate state:
\begin{equation}
reward(x_{t_i})=\alpha \cdot z_{\text{CLAP}}(x_{t_i})+(1-\alpha) \cdot z_{\text{PQ}}(x_{t_i}),
\end{equation}
where the weight $\alpha\in[0,1]$ can be adjusted to explicitly control the guidance process towards the direction of interest.
For example, if one desires a generation with high audio-text alignment, one can simply increase $\alpha$ to assign greater weight to the CLAP reward.
We note that this normalization and combination process is not limited to two rewards; any additional model with a distinct reward attribute can be incorporated and play as another axis of guidance for various applications. The effect of different weight values is further discussed in Section~\ref{4.3}.

\subsection{Audio-Text Alignment Evaluation with Audio Language Model}
Despite the growing interest in matching the generated audio to the text prompt, there are a limited number of relevant evaluation metrics. To gauge the performance of the proposed method from different perspectives, we introduce AQAScore, a novel audio-text alignment metric leveraging a pre-trained Audio Language Model named Audio Flamingo 3~\cite{af3}.
Specifically, we adopt the method used in VQAScore~\cite{vqascore}, where the image-text alignment is measured by asking a Vision Language Model a query \textit{“Does this image contain \{prompt\}?"} and calculating the probability of generating \textit{“yes"}.
For audio-text alignment evaluation, we feed Audio Flamingo 3 model an audio with a prompt \textit{“Does this audio contain \{prompt\}?"} and obtain the \textit{“yes"} probability.

%% file: tables/1_main_10.tex
\begin{table*}[t]
\centering
\caption{\textbf{Performance on the AudioCaps test set.} \emph{Scheme} indicates the reward strategy. $\mathsf{C}$ and $\mathsf{P}$ indicate the CLAP and PQ rewards, respectively. \emph{AQA} denotes AQAScore. $\alpha$ is the combination weight in SCORE. $\uparrow$ higher is better; $\downarrow$ lower is better.}
\small
\setlength{\tabcolsep}{5.5pt}
\resizebox{1.00\textwidth}{!}{\begin{tabular}{l l l|cc|cc|ccc}
\toprule
\multirow{2}{*}{Methods} & \multirow{2}{*}{Scheme} & \multirow{2}{*}{Setup} & \multicolumn{2}{c|}{Generation quality} & \multicolumn{2}{c|}{Text-Consistency} & \multicolumn{3}{c}{Audiobox-Aesthetics} \\
\cmidrule(lr){4-5}\cmidrule(lr){6-7}\cmidrule(lr){8-10}
& & & FD $\downarrow$ & IS $\uparrow$ & CLAP $\uparrow$ & AQA (\%) $\uparrow$ & PQ $\uparrow$ & CU $\uparrow$ & CE $\uparrow$ \\
\midrule
Naive sampling     & -                         & -                         &18.87{\scriptsize $\pm 0.37$} & 11.32{\scriptsize $\pm 0.24$} &0.65{\scriptsize $\pm 0.04$} & 90.05{\scriptsize $\pm 0.47$} &5.73{\scriptsize $\pm 0.24$} & 5.02{\scriptsize $\pm 0.21$} & 3.51{\scriptsize $\pm 0.12$} \\
\midrule
\multirow{4}{*}{Best-of-N}
  & Single reward           & $\mathsf{C}$                 &\underline{15.75}{\scriptsize $\pm 0.55$} & 11.63{\scriptsize $\pm 0.14$} &\underline{0.72}{\scriptsize $\pm 0.01$} & 91.57{\scriptsize $\pm 1.36$} &5.81{\scriptsize $\pm 0.01$} & 5.07{\scriptsize $\pm 0.02$} & 3.65{\scriptsize $\pm 0.03$} \\
  & Single reward           & $\mathsf{P}$                 &17.74{\scriptsize $\pm 0.78$} & 11.71{\scriptsize $\pm 0.22$} &0.67{\scriptsize $\pm 0.01$} & 89.49{\scriptsize $\pm 0.45$} & 6.32{\scriptsize $\pm 0.02$} & \underline{5.55}{\scriptsize $\pm 0.05$} & \underline{3.84}{\scriptsize $\pm 0.02$} \\
  & Rank aggregation & $\mathsf{C}+\mathsf{P}$   &16.89{\scriptsize $\pm 0.75$} & 11.85{\scriptsize $\pm 0.08$} &0.70{\scriptsize $\pm 0.02$} & 91.11{\scriptsize $\pm 0.75$} &6.21{\scriptsize $\pm 0.01$} & 5.45{\scriptsize $\pm 0.08$} & 3.81{\scriptsize $\pm 0.03$} \\
  & SCORE ($\alpha{=}0.50$) & $\mathsf{C}+\mathsf{P}$   &
17.24{\scriptsize $\pm 0.63$} & 11.86{\scriptsize $\pm 0.13$} &0.71{\scriptsize $\pm 0.01$} & 91.43{\scriptsize $\pm 0.24$} &6.21{\scriptsize $\pm 0.02$} & 5.44{\scriptsize $\pm 0.04$} &3.81{\scriptsize $\pm 0.03$} \\
\midrule
\multirow{4}{*}{Evosearch}
  & Rank aggregation & $\mathsf{C}+\mathsf{P}$    &15.92{\scriptsize $\pm 0.30$} & \underline{12.03}{\scriptsize $\pm 0.12$} &0.71{\scriptsize $\pm 0.00$} & 91.23{\scriptsize $\pm 0.52$} &6.16{\scriptsize $\pm 0.06$} & 5.35{\scriptsize $\pm 0.05$} & 3.73{\scriptsize $\pm 0.02$} \\
  & SCORE ($\alpha{=}0.50$) & $\mathsf{C}+\mathsf{P}$  &\underline{15.75}{\scriptsize $\pm 0.40$} & 11.93{\scriptsize $\pm 0.21$} &
\underline{0.72}{\scriptsize $\pm 0.00$} & \underline{91.63}{\scriptsize $\pm 0.53$} &\underline{6.34}{\scriptsize $\pm 0.02$} & 5.53{\scriptsize $\pm 0.01$} & 3.82{\scriptsize $\pm 0.01$} \\
  & SCORE ($\alpha{=}0.25$) & $\mathsf{C}+\mathsf{P}$  &16.54{\scriptsize $\pm 0.10$} & \textbf{12.04}{\scriptsize $\pm 0.30$} &0.69{\scriptsize $\pm 0.00$} & 90.61{\scriptsize $\pm 0.40$} &\textbf{6.47}{\scriptsize $\pm 0.02$} & \textbf{5.66}{\scriptsize $\pm 0.01$} & \textbf{3.86}{\scriptsize $\pm 0.01$} \\
  & SCORE ($\alpha{=}0.75$) & $\mathsf{C}+\mathsf{P}$  &\textbf{15.16}{\scriptsize $\pm 0.26$} & 11.86{\scriptsize $\pm 0.26$} & \textbf{0.74}{\scriptsize $\pm 0.00$} & \textbf{92.35}{\scriptsize $\pm 0.63$} & 6.09{\scriptsize $\pm 0.04$} & 5.29{\scriptsize $\pm 0.04$} & 3.70{\scriptsize $\pm 0.02$} \\
\bottomrule
\end{tabular}}
\label{table:main}
\end{table*}

%% file: figs/01_distribution_2.tex
\begin{figure}[t]
    \begin{subfigure}[t]{0.47\linewidth}
        \centering
        \includegraphics[width=\linewidth]{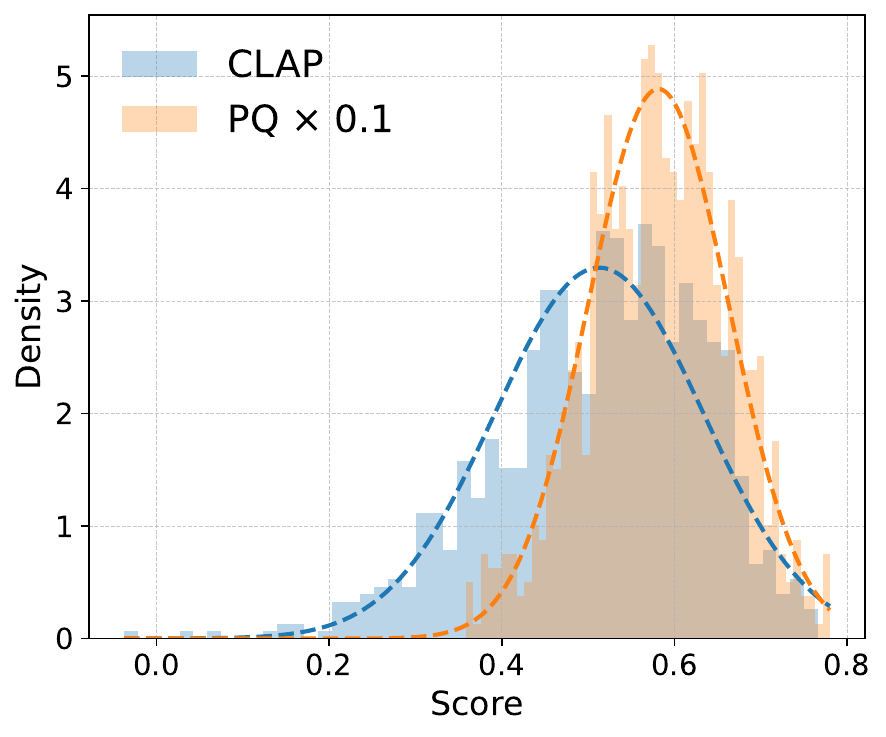}
        \caption{Raw distributions}
        \label{fig:boxplot_raw}
    \end{subfigure}
    \hfill
    \begin{subfigure}[t]{0.47\linewidth}
        \centering
        \includegraphics[width=\linewidth]{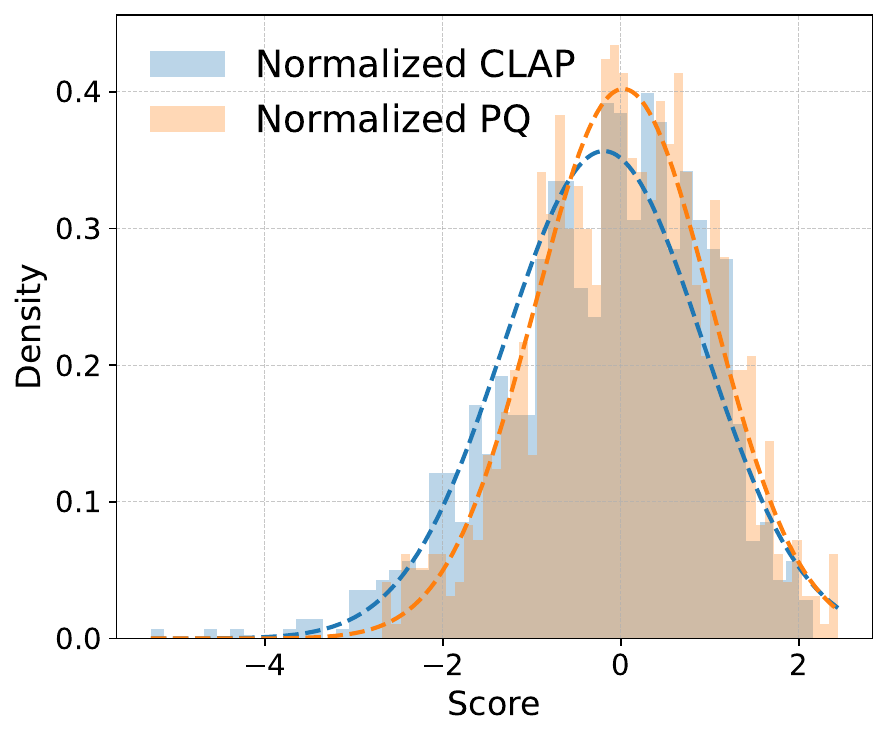}
        \caption{Normalized distributions}
        \label{fig:boxplot_z}
    \end{subfigure}
    \vspace{-2mm}
    \caption{\textbf{Comparison of reward score distributions.} (a) Raw distributions show large differences in mean and variance between CLAP and PQ×0.1. (b) After normalization (based on train-set statistics), both distributions 
    are aligned to zero mean and unit variance, enabling fair combination.}
    \vspace{-1mm}
    \label{fig:distribution}
\end{figure}

%% file: sec/3_experiment.tex
\section{Experiments}

\subsection{Experimental Setup}
\newpara{Implementation Details.}
We conduct experiments using EzAudio~\cite{hai2024ezaudio}, a T2A model.
For reward models, we utilize pretrained CLAP~\cite{clap} and Audiobox Aesthetics~\cite{audiobox_aesthetics} models.
We use the test split of AudioCaps~\cite{kim-etal-2019-audiocaps} dataset, consisting of 944 audio files with their textual descriptions.
The evaluation is conducted within two distinct sampling frameworks, BON and EvoSearch.
We report the average performance and standard deviation over three runs with distinct random seeds for reliability.
All experiments are conducted on NVIDIA RTX 4090 GPUs.

\newpara{Evaluation Metrics.}
Audio samples are evaluated with various metrics to gauge the performance in different aspects.
For generation quality, we use Frechet Distance (FD)~\cite{fd} and Inception Score (IS)~\cite{inceptionscore}.
In addition, three metrics proposed in Audiobox-Aesthetics~\cite{audiobox_aesthetics} are measured to further analyze the audio quality: Production Quality (PQ), Content Usage (CU), and Content Enjoyment (CE).
We do not report Production Complexity (PC) since evaluating the complexity of the generated audio is not our main objective.
We use CLAP similarity~\cite{clap} as well as the aforementioned AQAScore for audio-text alignment.

%% file: sec/4_results.tex
\section{Results}

\subsection{Main Results}
\newpara{Limitations on Single-Objective Guidance.}
We first compare the performance of different rewarding schemes with BON method. As shown in \Tref{table:main}, ITS utilizing a single reward model shows a clear trade-off across different quality attributes.
Compared to naive sampling, using the CLAP model as a reward improves the CLAP score by $10.8\%$ but increases the PQ score by only $1.4\%$.
On the other hand, using the PQ model as a single reward improves all Audiobox-Aesthetics metrics by approximately $10\%$, but marginally improves CLAP metric by $3\%$.
This clearly shows the over-reliance on a single reward model, and the issue is effectively addressed by utilizing multiple reward models.
Using rank aggregation and the proposed SCORE, the generation not only improves both CLAP by $9.2\%$ and PQ by $8.4\%$, but also clearly enhances all other metrics.
This demonstrates that “verifier hacking" can be effectively addressed with multiple reward models.

\newpara{Effectiveness of Standardized Composite Reward.}
The effect of the proposed SCORE is more pronounced in Evosearch framework.
As detailed in \Tref{table:main}, SCORE with equal weights on the two reward models ($\alpha=0.5$) improves CLAP by $10.8\%$ and PQ by $10.6\%$ over naive sampling.
Moreover, the improvement is demonstrated beyond the scope of the rewards, resulting in $10.2\%$ relative increase in CU and $8.8\%$ in CE.
We verify the advantage of our method against existing rank aggregation by consistently outperforming it across all evaluation metrics.
This is attributed to the utilization of full-precision scores, which provides more granular guidance than simple ordinal ranking.
The influence of SCORE is diminished in BON framework since the method can only be applied once at the last selection stage.
We note that SCORE is more effective in frameworks similar to EvoSearch because the iterative refinement of intermediate latents can provide strong positive guidance towards multiple aspects of audio.

\newpara{Evaluation with AQAScore.}
To evaluate the audio-text alignment without potential bias introduced by CLAP reward, we gauge the performance with the proposed AQAScore.
While AQAScore broadly follows the trend of CLAP, it reveals discrepancies. For example, BON with PQ reward attains a lower AQAScore than naive sampling despite a higher CLAP score, an evidence of verifier hacking.
Such observation provides an additional perspective on alignment evaluation, and we validate that AQAScore faithfully reflects the effectiveness of SCORE.

\subsection{Ablation Study}

\newpara{Effectiveness of Combination Weight.} \label{4.3}
We further investigate the effect of the combination weight $\alpha$ in the generation process, and demonstrate that it enables an intuitive control over inference towards desired attributes.
As shown in \Tref{table:main}, setting $\alpha=0.25$ to prioritize audio quality significantly boosts all quality-related metrics (PQ, CU, CE) at only a slight expense to textual alignment.
Conversely, prioritizing text alignment by setting $\alpha=0.75$ achieves the highest CLAP score among all methods, while the PQ score remains highly competitive, successfully avoiding the quality collapse observed in the single-objective case.
Finally, a balanced weight of $\alpha=0.5$ produces a faithful equilibrium between the two objectives, as demonstrated by its strong performance across all metrics.

\newpara{Effect of Number of Function Evaluations.}
To verify the scalability of the proposed method, we conduct experiments by applying SCORE to two different ITS methods with an incremental number of function evaluations (NFE).
As shown in \Fref{fig:population}, both methods show that the performance on CLAP and PQ metrics is proportional to the increase in NFE.
As indicated by the decreasing variance of the metrics, increasing computation makes the generation more robust to randomness and consistently produces favorable results.
Note that the population size $N$, the number of noise candidates in one inference, for EvoSearch is set lower than that of BON to match the NFE, and yet it shows a more stable increase in performance and mostly outperforms BON when scaled.

%% file: sec/5_conclusion.tex
\input{figs/02_population_2}

\section{Conclusion}

This paper pioneers the extension of Inference-Time Scaling to Text-to-Audio generation and presents a novel standardized composite reward framework named SCORE.
By normalizing and combining multiple reward signals, SCORE effectively balances audio–text alignment and perceptual quality while avoiding verifier hacking. Extensive experiments show consistent improvements across CLAP, PQ, and other evaluation metrics, with controllable trade-offs enabled through adjustable weighting. These results highlight SCORE as a principled and flexible approach for enhancing audio generation, setting a foundation for future work on multi-objective guidance in diffusion-based generative models.

%% file: figs/02_population_2.tex
\begin{figure}[t]
    \centering
    \begin{subfigure}[t]{0.48\linewidth}
        \centering
        \includegraphics[width=\linewidth]{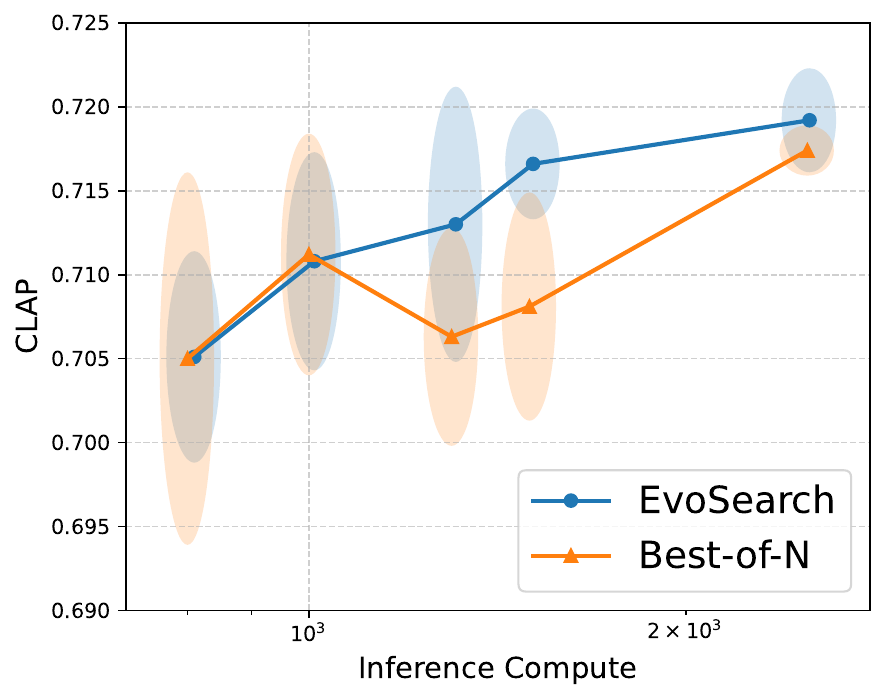}
        \caption{CLAP score}
        \label{fig:clap}
    \end{subfigure}
    \hfill
    \begin{subfigure}[t]{0.48\linewidth}
        \centering
        \includegraphics[width=\linewidth]{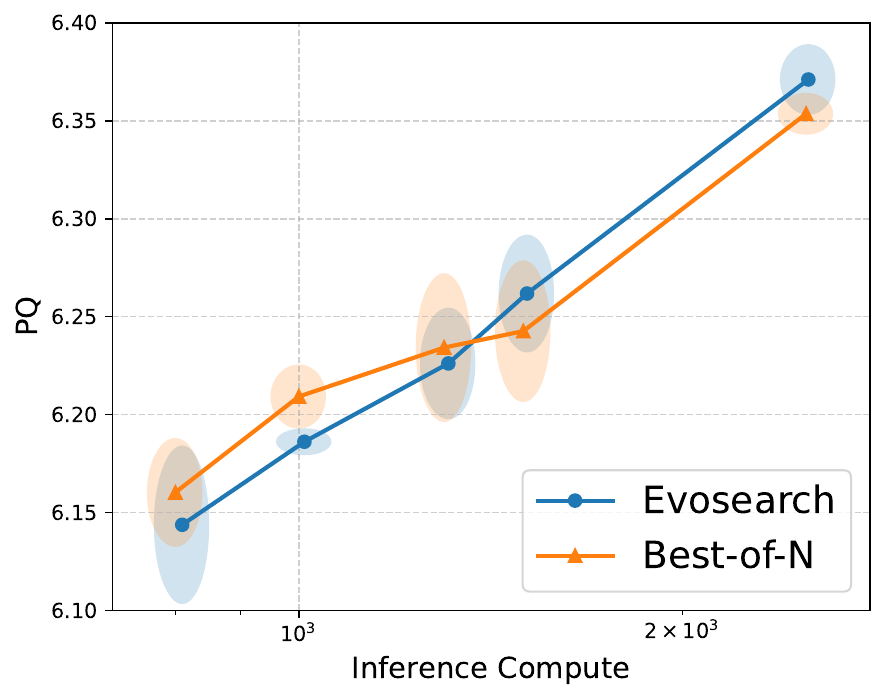}
        \caption{Audiobox-Aesthetics - PQ}
        \label{fig:pq}
    \end{subfigure}
    \vspace{-2mm}
    \caption{
        \textbf{Scaling performance of SCORE-guided EvoSearch and Best-of-N.}
        Larger population $N$ yield higher CLAP and PQ scores.
        The shaded ellipses represent the standard deviation across three random seeds.
    }
    \label{fig:population}
\end{figure}

%% file: main.bbl
\begin{thebibliography}{10}

\bibitem{audiogen}
Felix Kreuk, Gabriel Synnaeve, Adam Polyak, Uriel Singer, Alexandre Défossez, Jade Copet, Devi Parikh, Yaniv Taigman, and Yossi Adi,
\newblock ``Audiogen: Textually guided audio generation,''
\newblock in {\em Proc. ICLR}, 2023.

\bibitem{ho2020denoising}
Jonathan Ho, Ajay Jain, and Pieter Abbeel,
\newblock ``Denoising diffusion probabilistic models,''
\newblock in {\em Proc. NeurIPS}, 2020.

\bibitem{song2020denoising}
Jiaming Song, Chenlin Meng, and Stefano Ermon,
\newblock ``Denoising diffusion implicit models,''
\newblock in {\em Proc. ICLR}, 2021.

\bibitem{audioldm}
Haohe Liu, Zehua Chen, Yi~Yuan, Xinhao Mei, Xubo Liu, Danilo Mandic, Wenwu Wang, and Mark~D. Plumbley,
\newblock ``Audioldm: text-to-audio generation with latent diffusion models,''
\newblock in {\em Proc. ICML}, 2023.

\bibitem{audioldm2}
Haohe Liu, Yi~Yuan, Xubo Liu, Xinhao Mei, Qiuqiang Kong, Qiao Tian, Yuping Wang, Wenwu Wang, Yuxuan Wang, and Mark~D Plumbley,
\newblock ``Audioldm 2: Learning holistic audio generation with self-supervised pretraining,''
\newblock {\em IEEE/ACM Trans. on Audio, Speech, and Language Processing}, 2024.

\bibitem{evans2024fast}
Zach Evans, CJ~Carr, Josiah Taylor, Scott~H Hawley, and Jordi Pons,
\newblock ``Fast timing-conditioned latent audio diffusion,''
\newblock in {\em Proc. ICML}, 2024.

\bibitem{hai2024ezaudio}
Jiarui Hai, Yong Xu, Hao Zhang, Chenxing Li, Helin Wang, Mounya Elhilali, and Dong Yu,
\newblock ``Ezaudio: Enhancing text-to-audio generation with efficient diffusion transformer,''
\newblock in {\em Proc. Interspeech}, 2025.

\bibitem{jung2025voicedit}
Jaemin Jung, Junseok Ahn, Chaeyoung Jung, Tan~Dat Nguyen, Youngjoon Jang, and Joon~Son Chung,
\newblock ``Voicedit: Dual-condition diffusion transformer for environment-aware speech synthesis,''
\newblock in {\em Proc. ICASSP}, 2025.

\bibitem{huang2023make}
Rongjie Huang, Jiawei Huang, Dongchao Yang, Yi~Ren, Luping Liu, Mingze Li, Zhenhui Ye, Jinglin Liu, Xiang Yin, and Zhou Zhao,
\newblock ``Make-an-audio: Text-to-audio generation with prompt-enhanced diffusion models,''
\newblock in {\em Proc. ICML}, 2023.

\bibitem{ghosal2023text}
Deepanway Ghosal, Navonil Majumder, Ambuj Mehrish, and Soujanya Poria,
\newblock ``Text-to-audio generation using instruction guided latent diffusion model,''
\newblock in {\em Proc. ACM MM}, 2023.

\bibitem{its_diffusion}
Nanye Ma, Shangyuan Tong, Haolin Jia, Hexiang Hu, Yu-Chuan Su, Mingda Zhang, Xuan Yang, Yandong Li, Tommi Jaakkola, Xuhui Jia, et~al.,
\newblock ``Inference-time scaling for diffusion models beyond scaling denoising steps,''
\newblock {\em arXiv preprint arXiv:2501.09732}, 2025.

\bibitem{xie2025sana}
Enze Xie, Junsong Chen, Yuyang Zhao, Jincheng Yu, Ligeng Zhu, Chengyue Wu, Yujun Lin, Zhekai Zhang, Muyang Li, Junyu Chen, et~al.,
\newblock ``Sana 1.5: Efficient scaling of training-time and inference-time compute in linear diffusion transformer,''
\newblock in {\em Proc. ICML}, 2025.

\bibitem{lifshitzmulti}
Shalev Lifshitz, Sheila~A McIlraith, and Yilun Du,
\newblock ``Multi-agent verification: Scaling test-time compute with multiple verifiers,''
\newblock in {\em ICLR Workshop on Reasoning and Planning for Large Language Models}, 2025.

\bibitem{particlesampling_video}
Haolin Yang, Feilong Tang, Ming Hu, Qingyu Yin, Yulong Li, Yexin Liu, Zelin Peng, Peng Gao, Junjun He, Zongyuan Ge, et~al.,
\newblock ``Scalingnoise: Scaling inference-time search for generating infinite videos,''
\newblock {\em arXiv preprint arXiv:2503.16400}, 2025.

\bibitem{evosearch}
Haoran He, Jiajun Liang, Xintao Wang, Pengfei Wan, Di~Zhang, Kun Gai, and Ling Pan,
\newblock ``Scaling image and video generation via test-time evolutionary search,''
\newblock {\em arXiv preprint arXiv:2505.17618}, 2025.

\bibitem{particleLLM}
Isha Puri, Shivchander Sudalairaj, Guangxuan Xu, Kai Xu, and Akash Srivastava,
\newblock ``Rollout roulette: A probabilistic inference approach to inference-time scaling of llms using particle-based monte carlo methods,''
\newblock {\em arXiv preprint arXiv:2502.01618}, 2025.

\bibitem{brown2024large}
Bradley Brown, Jordan Juravsky, Ryan Ehrlich, Ronald Clark, Quoc~V Le, Christopher R{\'e}, and Azalia Mirhoseini,
\newblock ``Large language monkeys: Scaling inference compute with repeated sampling,''
\newblock {\em arXiv preprint arXiv:2407.21787}, 2024.

\bibitem{snell2024scaling}
Charlie Snell, Jaehoon Lee, Kelvin Xu, and Aviral Kumar,
\newblock ``Scaling llm test-time compute optimally can be more effective than scaling model parameters,''
\newblock in {\em Proc. ICLR}, 2025.

\bibitem{wu2024inference}
Yangzhen Wu, Zhiqing Sun, Shanda Li, Sean Welleck, and Yiming Yang,
\newblock ``Inference scaling laws: An empirical analysis of compute-optimal inference for problem-solving with language models,''
\newblock in {\em Proc. ICLR}, 2025.

\bibitem{reflectdit}
Shufan Li, Konstantinos Kallidromitis, Akash Gokul, Arsh Koneru, Yusuke Kato, Kazuki Kozuka, and Aditya Grover,
\newblock ``{Reflect-DiT: Inference-Time Scaling for Text-to-Image Diffusion Transformers via In-Context Reflection},''
\newblock {\em arXiv preprint arXiv:2503.12271}, 2025.

\bibitem{singhal2025general}
Raghav Singhal, Zachary Horvitz, Ryan Teehan, Mengye Ren, Zhou Yu, Kathleen McKeown, and Rajesh Ranganath,
\newblock ``A general framework for inference-time scaling and steering of diffusion models,''
\newblock in {\em Proc. ICML}, 2025.

\bibitem{liu2025video}
Fangfu Liu, Hanyang Wang, Yimo Cai, Kaiyan Zhang, Xiaohang Zhan, and Yueqi Duan,
\newblock ``Video-t1: Test-time scaling for video generation,''
\newblock in {\em Proc. ICCV}, 2025.

\bibitem{its_audiosr}
Yizhu Jin, Zhen Ye, Zeyue Tian, Haohe Liu, Qiuqiang Kong, Yike Guo, and Wei Xue,
\newblock ``{Inference-time Scaling for Diffusion-based Audio Super-resolution},''
\newblock {\em arXiv preprint arXiv:2508.02391}, 2025.

\bibitem{af3}
Arushi Goel, Sreyan Ghosh, Jaehyeon Kim, Sonal Kumar, Zhifeng Kong, Sang-gil Lee, Chao-Han~Huck Yang, Ramani Duraiswami, Dinesh Manocha, Rafael Valle, et~al.,
\newblock ``Audio flamingo 3: Advancing audio intelligence with fully open large audio language models,''
\newblock {\em arXiv preprint arXiv:2507.08128}, 2025.

\bibitem{vqascore}
Zhiqiu Lin, Deepak Pathak, Baiqi Li, Jiayao Li, Xide Xia, Graham Neubig, Pengchuan Zhang, and Deva Ramanan,
\newblock ``Evaluating text-to-visual generation with\&nbsp;image-to-text generation,''
\newblock in {\em Proc. ECCV}, 2024.

\bibitem{clap}
Yusong Wu, Ke~Chen, Tianyu Zhang, Yuchen Hui, Taylor Berg-Kirkpatrick, and Shlomo Dubnov,
\newblock ``Large-scale contrastive language-audio pretraining with feature fusion and keyword-to-caption augmentation,''
\newblock in {\em Proc. ICASSP}, 2023.

\bibitem{audiobox_aesthetics}
Andros Tjandra, Yi-Chiao Wu, Baishan Guo, John Hoffman, Brian Ellis, Apoorv Vyas, Bowen Shi, Sanyuan Chen, Matt Le, Nick Zacharov, et~al.,
\newblock ``Meta audiobox aesthetics: Unified automatic quality assessment for speech, music, and sound,''
\newblock {\em arXiv preprint arXiv:2502.05139}, 2025.

\bibitem{jin2025inference}
Yizhu Jin, Zhen Ye, Zeyue Tian, Haohe Liu, Qiuqiang Kong, Yike Guo, and Wei Xue,
\newblock ``Inference-time scaling for diffusion-based audio super-resolution,''
\newblock {\em arXiv preprint arXiv:2508.02391}, 2025.

\bibitem{eyring2024reno}
Luca Eyring, Shyamgopal Karthik, Karsten Roth, Alexey Dosovitskiy, and Zeynep Akata,
\newblock ``Reno: Enhancing one-step text-to-image models through reward-based noise optimization,''
\newblock in {\em Proc. NeurIPS}, 2024.

\bibitem{kim2025test}
Sunwoo Kim, Minkyu Kim, and Dongmin Park,
\newblock ``Test-time alignment of diffusion models without reward over-optimization,''
\newblock in {\em Proc. ICLR}, 2025.

\bibitem{kim-etal-2019-audiocaps}
Chris~Dongjoo Kim, Byeongchang Kim, Hyunmin Lee, and Gunhee Kim,
\newblock ``{A}udio{C}aps: Generating captions for audios in the wild,''
\newblock in {\em Proc. NAACL}, 2019.

\bibitem{fd}
Thomas Eiter, Heikki Mannila, et~al.,
\newblock ``Computing discrete fr{\'e}chet distance,''
\newblock 1994.

\bibitem{inceptionscore}
Shane Barratt and Rishi Sharma,
\newblock ``A note on the inception score,''
\newblock {\em arXiv preprint arXiv:1801.01973}, 2018.

\end{thebibliography}
